\documentclass{cup-hpl}

\usepackage{graphicx}% 
\usepackage{hyperref}% 
\begin{document}

\shorttitle{Magnetization dynamics driven by structured lasers.}                                   
\shortauthor{L. Sánchez-Tejerina et al.}

\title{All-optical non-linear chiral ultrafast magnetization dynamics driven by circularly polarized magnetic fields}

\author[1,2]{Luis Sánchez-Tejerina\corresp{\email{luis.stsj@usal.es}}}
\author[1]{Rodrigo Martín-Hernández}
\author[3]{Rocío Yanes}
\author[1]{Luis Plaja}
\author[3]{Luis López-Díaz}
\author[1]{Carlos Hernández-García}

\address[1]{Grupo de Investigación en Aplicaciones del Láser y Fotónica, Departamento de Física Aplicada, Universidad de Salamanca, E-37008, Salamanca, Spain.}
\address[2]{Present address: Departamento de Electricidad y Electrónica, Universidad de Valladolid, 47011 Valladolid, Spain}
\address[3]{Departamento de Física Aplicada, Universidad de Salamanca, E-37008, Salamanca, Spain}

\begin{abstract} %Ultrafast 
Ultrafast laser pulses provide unique tools to manipulate magnetization dynamics at femtosecond timescales, where the interaction of the electric field usually dominates over the magnetic field. Recent proposals using structured laser beams have demonstrated the possibility to produce regions where intense oscillating magnetic fields are isolated from the electric field. In these conditions, we show that technologically feasible Tesla-scale circularly polarized high-frequency magnetic fields induce purely precessional nonlinear magnetization dynamics. This fundamental result not only opens an avenue in the study of laser-induced ultrafast magnetization dynamics, but also sustains technological implications as a route to promote all-optical non-thermal magnetization dynamics both at shorter timescales---towards the sub-femtosecond regime--- and at THz frequencies.
% 112 palabras. Máximo 150 palabras
\end{abstract}
 
\keywords{Ultrafast dynamics, Non-linear dynamics, Chiral behavior}

\maketitle

\section{Introduction}
The pioneering work on ultrafast demagnetization in Ni \cite{Beaurepaire:1996} paved the way towards a large number of theoretical and experimental studies on magnetization dynamics at the femtosecond (fs) time scales induced by ultrashort laser pulses \cite{Stanciu:2007,Schmidt:2010,Tengdin:2018,Tesarova:13,Lingos:2015,Choi:17,Stupakiewicz:2017,Davies:2020,Zhang:00b,Koopmans:2005,Rudolf:2012,Krieger:2015,Bonetti:2016,Bierbrauer:2017,Dewhurst:2018,Siegrist:2019,Tengdin:2020,Hofherr:2020,Scheid:2021,Chekhov:2021,Kimel:2005,Hamamera:2022,Watzel:2016,Zhang:2016b}. In these studies the dynamics is mediated primarily by the electric field (E-field), which can excite non-equilibrium states \cite{Tesarova:13,Lingos:2015,Choi:17,Stupakiewicz:2017,Davies:2020}, demagnetize the sample, \cite{Beaurepaire:1996,Bonetti:2016,Bierbrauer:2017,Dewhurst:2018,Siegrist:2019,Tengdin:2020,Hofherr:2020,Scheid:2021,Chekhov:2021} generate localized charge currents \cite{Watzel:2016,Zhang:2016b}, or induce the inverse Faraday effect \cite{Kimel:2005,Hamamera:2022}. 
While most of the techniques are mediated mainly by the E-field, other techniques, such as the excitation of phononic modes \cite{Stupakiewicz:2021}, have recently provided routes for non-thermal magnetization manipulation.

An appealing alternative to induce coherent magnetization dynamics consists on the use of magnetic fields (B-field). The role of the B-field in ultrafast magnetization dynamics has been extensively studied, specially in the regime of linear response to THz fields \cite{Tudosa:2004,Wienholdt:2012,Vicario:2013,Bocklage:2016,Blank:2021,Choi:2021}. At this picosecond time scale, few Tesla (T) are required to introduce small deflections from the equilibrium magnetization direction, while tens of T are needed for achieving complete switching. Higher driving frequencies, that could break into the femtosecond timescale, would require very high B-field amplitudes.Although intense magnetic fields can be achieved, for example, using plasmonic antennas\cite{Yang:2022}, in such regime, the associated E-field would potentially demagnetize the sample\cite{Shalaby:2018} or even damage it. Besides, although substantial advances have been made towards the generation of electromagnetic fields in the range of THz ($0.1$ to $30\;\mathrm{THz}$), their intensity is still small as compared to the infrared case \cite{Kampfrath:2013,Salen:2019,Seifert:2022}. 

In this work we introduce an appealing alternative to drive magnetization dynamics at the sub-picosecond timescale, by using isolated ultrafast intense B-fields.
Recent developments in structured laser sources have demonstrated the possibility to spatially decouple the B-field from the E-field of an ultrafast laser pulse. For instance, azimuthally-polarized laser beams present a longitudinal B-field at the beam axis, where the E-field is zero \cite{zhan_cylindrical_2009}. Depending on the laser beam parameters, the contrast between the B-field and E-field can be adjusted, so to design a local region in which the B-field can be considered to be isolated \cite{Blanco:2019}. In such region, the stochastic processes driven by the E-field could be avoided, and the coherent precession induced by the B-field can be exploited. Indeed, azimuthally-polarized laser beams have been shown to induce isolated mili-Tesla static B-fields \cite{Guclu:2016}, with applications in nanoscale magnetic excitations and photoinduced force microscopy \cite{Zeng:2018, Zeng:2018_2}. More recently, ultrafast time-resolved magnetic circular dichroism has been proposed \cite{Cao2022}. In addition, theoretical proposals \cite{Blanco:2019,Sederberg:2020} and experiments \cite{Sederberg:2020b, Jana:2021} have raised the possibility to generate isolated Tesla-scale fs magnetic fields by the induction of large oscillating currents through azimuthally polarized fs laser beams. 

Our theoretical study unveils the non-linear, chiral, precessional magnetization response of a standard ferromagnet to a Tesla scale circularly polarized ultrafast magnetic field whose polarization plane contains the initial equilibrium magnetization. First, we show in section~\ref{sec:2} the feasibility to use state-of-the-art structured laser beams to create a macroscopic region in which such B-fields are found to be isolated from the E-field by particle-in-cell (PIC) simulations. Then, we present our micromagnetic ($\mu$Mag) simulations for moderate fields in section~\ref{sec:3} showing the presence of measurable magnetization dynamics in CoFeB when a circularly polarized $10$ ps B-field pulse of $10\;\mathrm{T}$ and central frequency $30\;\mathrm{THz}$ is applied. Additionally, we compare the dynamics triggered by a B-field with linear polarization, circular polarization with the polarization plane perpendicular to the equilibrium magnetization, and circular polarization with the polarization plane parallel to the equilibrium magnetization. Measurable magnetization dynamics are found in the later case. In section~\ref{sec:4}, we provide for a complete analytical model to describe such dynamics, and compare it with full $\mu$Mag simulations. This model allows us to predict the complete magnetization switching by using 1 ps, $275\;\mathrm{T}$, $60\;\mathrm{THz}$, B-field pulses, verifed by full $\mu$Mag simulations. Finally, section~\ref{sec:5} summarizes the main conclusions of the work and gives some perspectives on possible implications in the field.

\section{Spatially isolated circularly polarized B-fields out of structured laser beams}\label{sec:2}

In order to study the interaction of an isolated, circularly polarized B-field with a standard ferromagnet (CoFeB), we consider a B-field, $\mathbf{B}$, oscillating in the $xz$ plane (see Fig. 1(a)) given by 
\begin{equation}
    \mathbf{B}\left(t\right)=\mathbf{b}\left(t\right)e^{i\omega t} +\mathbf{b}^*\left(t\right)e^{-i\omega t}
\end{equation}
\begin{equation}
    \mathbf{b}\left(t\right)=\frac{B_0}{2}F\left(t\right)\left(\cos\theta_0 \hat{u}_x + \sin\theta_0 e^{i\phi_0}\hat{u}_z\right),
\end{equation}
where $\omega$ is the central angular frequency, ($\omega=2\pi f$), $B_0$ is the amplitude, and $\theta_0$ and $\phi_0$ define the relative amplitude and phase between the $x$ and $z$ components, respectively. $F\left(t\right)$ is the field envelope, given by $F(t)=\sin^2(\pi t/T_p)$ for $0 \leq t \leq T_p$, with $T_p=3/8 t_p$ its full duration, $t_p$ being the full-width-at-half-maximum (FWHM) pulse duration in intensity. A right-handed---RCP--- (left-handed---LCP---) circularly polarized B-field in the $xz$ plane corresponds to $\phi_0=\pi/2$ ($\phi_0=-\pi/2$) and $\theta_0=\pi/4$, while a linearly polarized B-field corresponds to $\phi_0=0$ or $\pi$. 

\begin{figure}
\begin{center}
\includegraphics[width=\columnwidth]{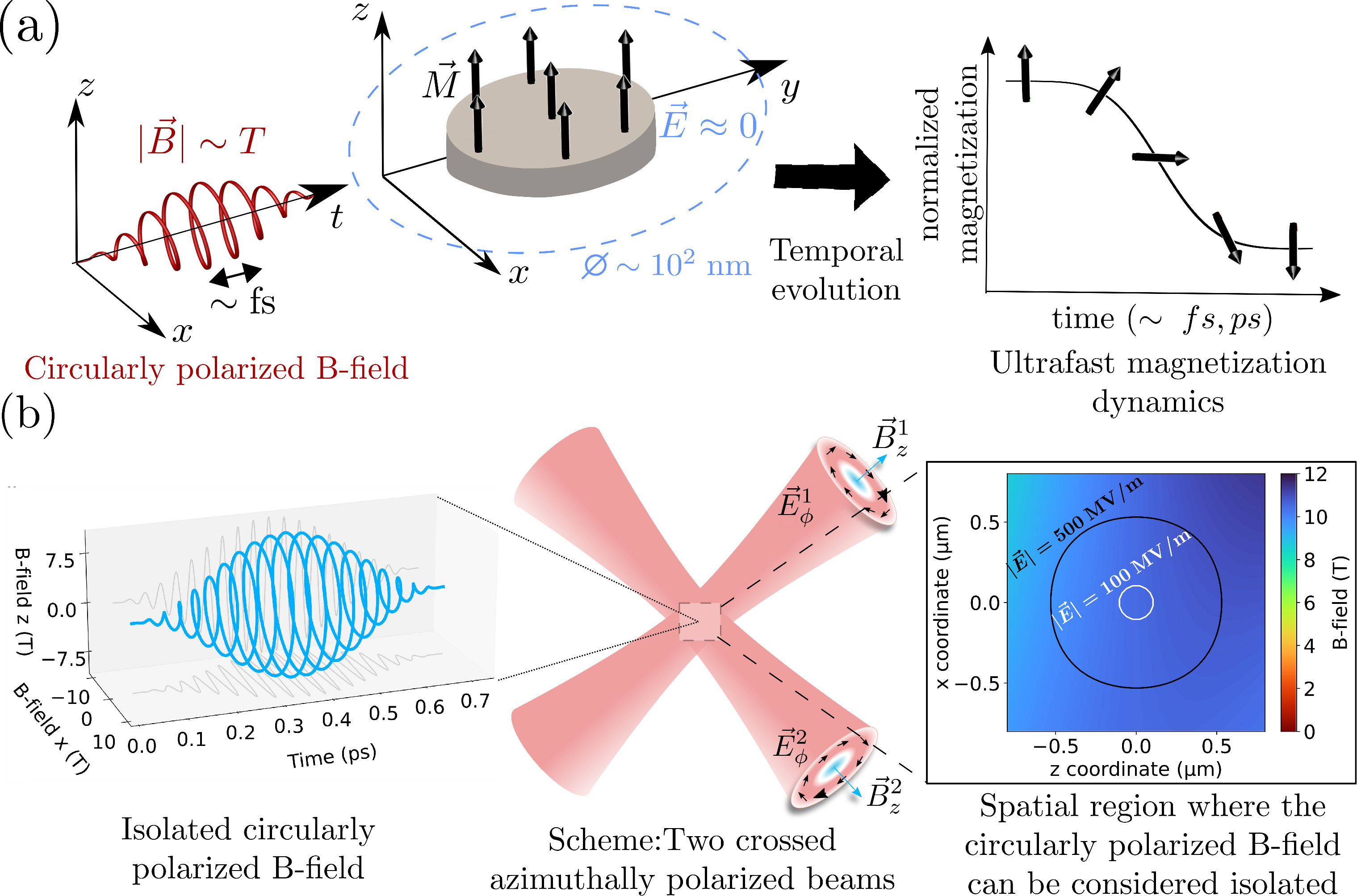}
\caption{%\label{fig:s1}
a) Sketch of the system under consideration. A circularly polarized magnetic field illuminates a magnetic sample whose dimensions are smaller than the region for which the E-field can be considered negligible. This field can trigger ultrafast magnetization dynamics. b) Two crossed azimuthally polarized beams of 30 THz and peak intensity 2.1$\times 10^{13}$ W/cm$^2$ define a spatial region of $\simeq 100$ nm in which the E-field is lower than 100 MV/m, as depicted in panel. In such region, a constant B-field of amplitude 10.5 T and central frequency 30 THz is found.}
\end{center}\vspace*{0cm}
\end{figure}

In our simulations, we do not include any E-field coupling, as the B-field is assumed to be isolated. Such assumption is valid for CoFeB in spatial regions where the E-field is lower than $100\;\mathrm{MV/m}$, for which the demagnetization has been predicted to be less than $7\%$ \cite{Bonetti:2016,Hudl:2019}. The conditions for which an intense circularly polarized B-field can be found spatially isolated from the E-field can be obtained by using two crossed azimuthally polarized laser beams, as sketched in Fig. 1(b). We have performed PIC simulations using the OSIRIS 3D PIC code \cite{Fonseca2003, Fonseca2008, Fonseca2013}, in order to show how such isolated B-fields can be achieved with the state-of-the-art ultrafast laser technology. We have considered two orthogonal azimuthally polarized laser beams with waist $w_0=3.125\lambda=31.25$ $\mu$m, a central wavelength of $\lambda=10$ $\mu$m (30 THz), and E-field amplitude of 12.5 GV/m (peak intensity of 2.1$\times 10^{13}$ W/cm$^2$) at their radius of maximum intensity, $w_0/\sqrt{2}$. The temporal envelope is modeled as a sin$^2$ function of 88.8 fs FWHM. Due to computational limitations the temporal envelope is much shorter than those considered in the $\mu$Mag simulations presented in this work, which lies in the ps regime. However, we do not foresee any deviation in the results presented if longer pulses with similar amplitudes are considered.

In Fig. 1(b) we also show the spatial distribution of the B-field (color background) and the E-field (contour lines) at overlapping region. We have highlighted the region in which the E-field is lower than $100\;\mathrm{MV/m}$, and thus the E-field can be neglected against the B-field. Thus, we can define a region of radius $\simeq 100$ nm in which the B-field exhibits a constant amplitude of 10.5 T and the E-field is maintained below $100$ MV/m. Though the use of additional currents, like in Refs. \cite{Blanco:2019, Sederberg:2020}, could enhance the B-field amplitude, our simulations demonstrate that moderately intense laser beams can already reach the B-field amplitudes required to observe the non-linear magnetization dynamics described below.

\section{Nonlinear magnetization response to ultrafast B-fields}\label{sec:3}

The interaction between the oscillating B-field and the magnetization is given by the Landau-Lifshitz-Gilbert (LLG) equation \cite{Wegrowe:2012,Vicario:2013}
\begin{equation}\label{eq:LLGcompl}
	\left(1+\alpha^2\right)\frac{d\mathbf{m}}{dt}=-\gamma\mathbf{m}\times\mathbf{B}_{eff}-\alpha\mathbf{m}\times\left(\mathbf{m}\times\mathbf{B}_{eff}\right)
\end{equation}
where $\mathbf{m}$ is the normalized magnetization where both spatial and temporal dependencies are implicitly assumed, $\alpha$ is the Gilbert damping parameter, and $\mathbf{B}_{eff}$ is the effective magnetic field.  We have performed $\mu$Mag simulations using the well-known software MuMax$^3$ \cite{mumax} to solve the LLG equation. The system under study is sketched in Fig. 1(a), where we consider a circular nanodot with $1\;\mathrm{nm}$ thickness and $64\;\mathrm{nm}$ diameter discretized into $1\;\mathrm{nm}$ cubic cells. The material parameters correspond to CoFeB grown over a heavy metal layer: inhomogeneous exchange parameter $A=19\;\mathrm{pJ/m}$, saturation magnetization $M_S=1\;\mathrm{MA/m}$, perpendicular uniaxial anisotropy (i.e. the anisotropy field is directed along the $z$ direction) parameter $K_u=800\;\mathrm{kJ/m^3}$, Dzyaloshinkii-Moriya interaction (DMI) $D=1.8\;\mathrm{mJ/m^2}$ and Gilbert damping $\alpha=0.015$. 

In Fig. 2(a) we show the in-plane magnetization dynamics (perpendicular to the equilibrium configuration, $m_z=1$) induced by RCP and LCP B-fields lying in the $xz$ plane. Note that the equilibrium magnetization lies in the polarization plane. In both cases, $B_0=10\;\mathrm{T}$, $f=30\;\mathrm{THz}$, and $t_p=10\;\mathrm{ps}$. We can observe a magnetization precession around the $z$ axis triggered by a non-linear chiral response to the B-field. While the RCP B-field induces a measurable negative $x$ component, the LCP leads to a positive one. After the pulse, the precession dynamics is dominated by the anisotropy field, and the system starts to precess around the z-axis. Note that the broad trace is due to the subsequent magnetization oscillations during the interaction with the pulse. 

The non-linear mechanism underlying such behavior can be understood as follows (see bottom part of Fig. 2(a)). At an initial time $t=0$, in which $\mathbf{m}$ (black arrow) lies in the polarization plane of the circularly polarized B-field (red arrow), being perpendicular to it, a transverse torque $\mathbf{\tau}$ (green arrow) drives $\mathbf{m}$ out-of-plane from this initial position. During the next quarter-period, $\mathbf{\tau}$ decreases and rotates, inducing a precession of $\mathbf{m}$ around its initial axis. For the second quarter-period, $\mathbf{\tau}$ increases again keeping its rotation but, at $t=T/2$, it reverses its rotation direction, thus sweeping only half of the plane perpendicular to $\mathbf{m}$. As a result, along a whole period, the torque component perpendicular to the polarization plane averages to zero, while  a residual contribution along the intersection of the polarization plane and the plane perpendicular to $\mathbf{m}$ remains. With long lasting multicycle laser pulses it is then possible to accumulate the small torque along the polar coordinate on the polarization plane, $\theta$, so as to promote the system to a targeted non-equilibrium state. This is reflected in Fig. 2(a), where the magnetization components $m_x$ and $m_y$ are non-zero at the end of the pulse, and therefore the magnetization is not aligned along the anisotropy direction, $z$. A more detailed scheme of the non-linear mechanism is displayed in Supplementary Videos 1 and 2 for both, RCP and LCP B-fields, revealing the chiral nature of the reported effect.

\begin{figure}
\includegraphics[width=\columnwidth]{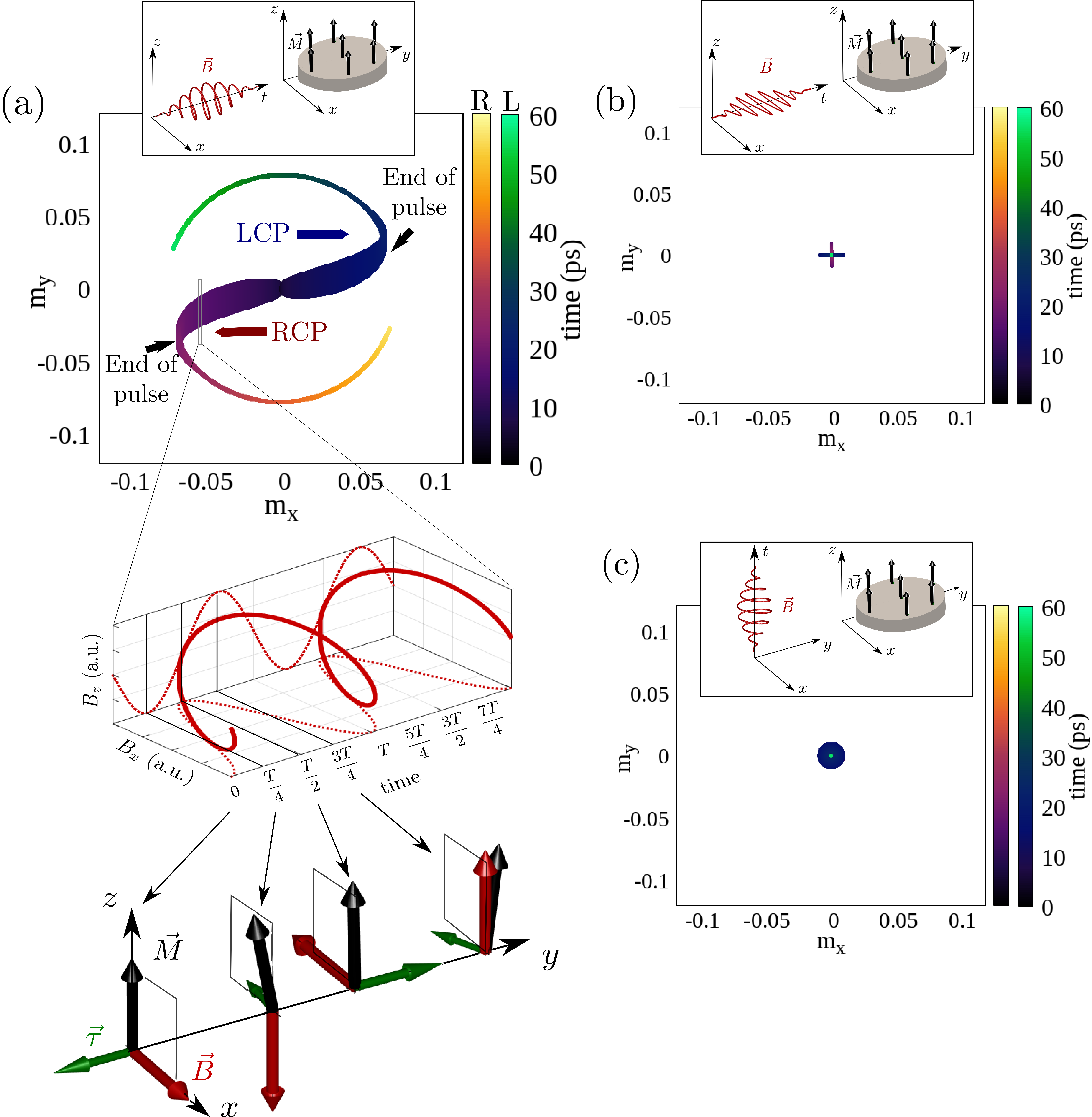}
\caption{%\label{fig:1}
Micromagnetic simulation results of the temporal evolution (color code) of the magnetization components ($m_x$, $m_y$) of CoFeB excited by B-fields with different polarization states. a) RCP (yellowish color scale) and LCP (greenish color scale) B-fields ($B_0=10\;\mathrm{T}$, $f=30$ THz, $t_p=10$ ps). The RCP (LCP) B-field induces a measurable negative (positive) $m_x$ component. In both cases the anisotropy field induces a precession of $\mathbf{m}$ around the equilibrium configuration. The bottom part sketches the mechanism during a B-field period of constant amplitude. The B-field (red), magnetization (black) and torque (green) vector representations at four different times reveal the magnetization dynamics mechanism over one period. b) Linear polarization along x (yellowish trace) or y (greenish trace). c) Circular polarization perpendicular to the equilibrium magnetization with RCP (yellowish trace) and LCP (greenish trace) helicities.}
\end{figure}

To highlight the importance of the polarization state and orientation to get the non-linear response, Figs. 2(b) and 2(c) depict the temporal evolution of the magnetization components ($m_x$, $m_y$) obtained from full micromagnetic simulations for a linearly polarized B-field, being perpendicular to the equilibrium magnetization, and for a circularly polarized B-field (either RCP or LCP), where the polarization plane is perpendicular to the equilibrium magnetization. Whereas in both cases a small magnetization deflection is observed, the net torque exerted by the field on the magnetization over a period is null, and the magnetization recovers its equilibrium state after the B-field pulse. In addition, at frequencies larger than few tens of THz the response is not enough to promote significant change on the magnetization even for a B-field as high as $B=10\;\mathrm{T}$. Consequently, in the cases presented in Figs. 2(b) and 2(c), the response is completely linear and the magnetization comes back to the initial configuration at the end of the B-field pulse. Nonetheless, for a circularly polarized B-field (either RCP or LCP) with the equilibrium magnetization lying in the polarization plane, the non-linear chiral phenomenon described above triggers the magnetization out of equilibrium as shown Fig. 2(a). This dragging, being a non-linear effect, is sensitive to the B-field envelope and does not cancel out at the end of the pulse.

\section{Analytical model}\label{sec:4}
To give insight into the nonlinear mechanism introduced in previous section and sketched in Fig. 2(a), we derive an approximated analytical model. The exchange field is not included in the model because we assume that the sample remains uniformly magnetized. Besides, we neglect the anisotropy and DMI fields---which are small if compared to the external one--- and the damping term. Similar assumptions has been proven reasonable at this time scale in previous studies\cite{Vicario:2013}. With these approximations in Eq. (\ref{eq:LLGcompl}), the magnetization dynamics out of the polarization plane reads as
\begin{equation}
    \frac{dm_y}{dt}\mathbf{u}_y=-\gamma'\mathbf{m}_\parallel\times\mathbf{B} ,
\end{equation}\label{eq:myDynamics}
$\mathbf{m}_\parallel$ being the magnetization in the polarization plane and $\gamma'=\gamma/\left(1+\alpha^2\right)$. Considering the initial magnetization in the $z$ direction, $m_y$ at any time $t$ is given by
\begin{equation}\label{eq:intmy}
	m_y\left(t\right)\mathbf{u}_y=-\gamma'\int_0^t\mathbf{m}_\parallel\left(\tau\right)\times\mathbf{B}\; d\tau.
\end{equation}
The cartesian components of the magnetization can be decomposed at each point in its Fourier components,
\begin{equation}\label{eq:Fourier}
    {m}_j\left(t\right)=\sum_q m_q^j\left(t\right)e^{iq\omega t} \quad\quad\quad j=\{ x,y,z\}.
\end{equation}
Using Eqs. (\ref{eq:intmy}) and (\ref{eq:Fourier}) in the simplified LLG equation, we obtain \vspace*{0.2cm}\\
$\displaystyle\sum_q \left(\frac{d\mathbf{m}_q^\parallel\left(t\right)}{dt}+iq\omega\mathbf{m}_q^\parallel\left(t\right)\right)e^{iq\omega t} =$\\
\hspace*{0.3cm} $+\gamma'{^2}\left[\displaystyle\sum_q\int_0^t\mathbf{m}_{q-1}^\parallel\left(\tau\right)\times\mathbf{b}\left(\tau\right)e^{iq\omega \tau}d\tau\right]\times\mathbf{B}\left(t\right)+$
\begin{equation}\label{eq:Complete}
+\gamma'{^2}\left[\displaystyle\sum_q\int_0^t\mathbf{m}_{q+1}^\parallel\left(\tau\right)\times\mathbf{b}^*\left(\tau\right) e^{iq\omega \tau} d\tau\right]\times\mathbf{B}\left(t\right).
\end{equation}
Assuming that the magnetization components in the polarization plane, $\mathbf{m}_{q\pm 1}^\parallel$, and the B-field envelope, $\mathbf{b(t)}$, evolve slowly, considering $\mathbf{b}\left(0\right)=0$, and selecting only the slowly varying terms ($q=0$), Eq. (\ref{eq:Complete}) transforms into
\begin{equation}\label{eq:Slowly}
    \frac{d\mathbf{m}_0^\parallel\left(t\right)}{dt} = -\frac{2i\gamma'{^2}}{\omega}\mathbf{m}_0^\parallel\left(t\right)\times\left(\mathbf{b}\left(t\right)\times\mathbf{b}^*\left(t\right)\right).
\end{equation}
It is well known that the effective field dependence on the magnetization can lead to non-linear effects \cite{Bertotti:2002,dAquino:2009,Hudl:2019}. However, it must be noticed that, differently from those cases, the described effect is non-linear on the external field, not on the effective field. Moreover, it is proportional to the gyromagnetic ratio and the inverse of the frequency, being equivalent to a drift magnetic field $\vec{B}_d$ given by,
\begin{equation}\label{eq:DriftField}
    \mathbf{B}_d = \frac{\gamma'}{2\omega}\sin\phi_0\left(\mathbf{B}_x\times\mathbf{B}_z\right).
\end{equation}
Using this definition, Eq. (\ref{eq:Slowly}) describes the slowly varying LLG dynamics in terms of the drift field, $\mathbf{B}_d$. Eqs. (\ref{eq:Slowly}) and (\ref{eq:DriftField}) constitute the main contribution of the present work, as they reveal a second-order dependency of the magnetization dynamics with the external B-field. From Eq. (\ref{eq:DriftField}) we can already infer that $\mathbf{B}_d$ is maximal for circular polarization, decreases with the ellipticity, and is zero for a linearly polarized B-field ($\phi_0=0$ or $\pi$). Note also the chiral nature of the presented mechanism, as the direction of $\mathbf{B}_d$ is helicity dependent. Finally, we stress the purely precessional nature of $\mathbf{B}_d$---being linear with the gyromagnetic ratio---and its inverse proportionality with the driving frequency. It is worth noting that a small misalignment of the azimuthally polarized laser beams would convert the circularly polarized magnetic field into elliptically polarized magnetic field, and/or would introduce a small angle between the initial magnetization and the polarization plane. Nonetheless, it is possible to decompose the total magnetic field in a circularly polarized magnetic field in the $xz$ plane and a linearly polarized magnetic field in the $y$ direction. However, this later component would not affect the slow dynamics presented here.

We now analyze the dependency of the magnetization dynamics on the B-field, both with the analytical model represented by Eq. (\ref{eq:Slowly}), and the full micromagnetic simulations, where all the interactions on the effective field, as well as the damping, are included. To highlight the accuracy of our model based on the equivalent drift field, we compare the total rotation of the magnetization from our simulations with the magnetization rotation induced by the drift B-field, $\mathbf{B}_d$, which can be computed as
\begin{equation}\label{eq:maxrot}
    \Delta\theta = \gamma'\left[\frac{\gamma'}{2\omega}\sin\phi_0\left(B_x B_z\right)\right]t_p.
\end{equation}

Fig. 3, presents the induced magnetization rotation as derived from the analytical model (solid lines) and the micromagnetic simulations (dots). The excellent agreement allows us to validate our model and demonstrate the reported non-linear chiral effect. First, Fig. 3(a) shows the total rotation of the magnetization as a function of the polarization state (characterized by $\phi_0$) of an external B-field of $t_p=3$ ps, for amplitudes of 60 T (blue) , 100 T (red) and 140 T (black). Our simulations confirm no rotation for a linearly polarized B-field, and a maximum rotation for circular polarization. The chiral character of the phenomenon is also evidenced.

Fig. 3(b) depicts the inverse dependency of the magnetization rotation with the B-field frequency. This frequency scaling suggests that the non-linear induced rotation is particularly relevant for external B-fields at THz frequencies. However, note that the linear dynamics (with the external field) would also contribute at those frequencies. Fig. 3(c) shows the second-order scaling of the magnetization dynamics with the external B-field amplitude for central frequencies of 250 THz (blue), 100 THz (red) and 50 THz (black). As expected, the total rotation increases with the B-field amplitude, being already measurable at tens of T. Finally, Fig. 3(d) depicts the total rotation of the magnetization for a B-field pulse of frequency $50\;\mathrm{THz}$ as a function of the pulse duration, $t_p$. This latter result confirms that the non-linear chiral effect presented in this work is cumulative in time, as predicted from Eq. (\ref{eq:maxrot}). 

\begin{figure}
\begin{center}
\includegraphics[width=\columnwidth]{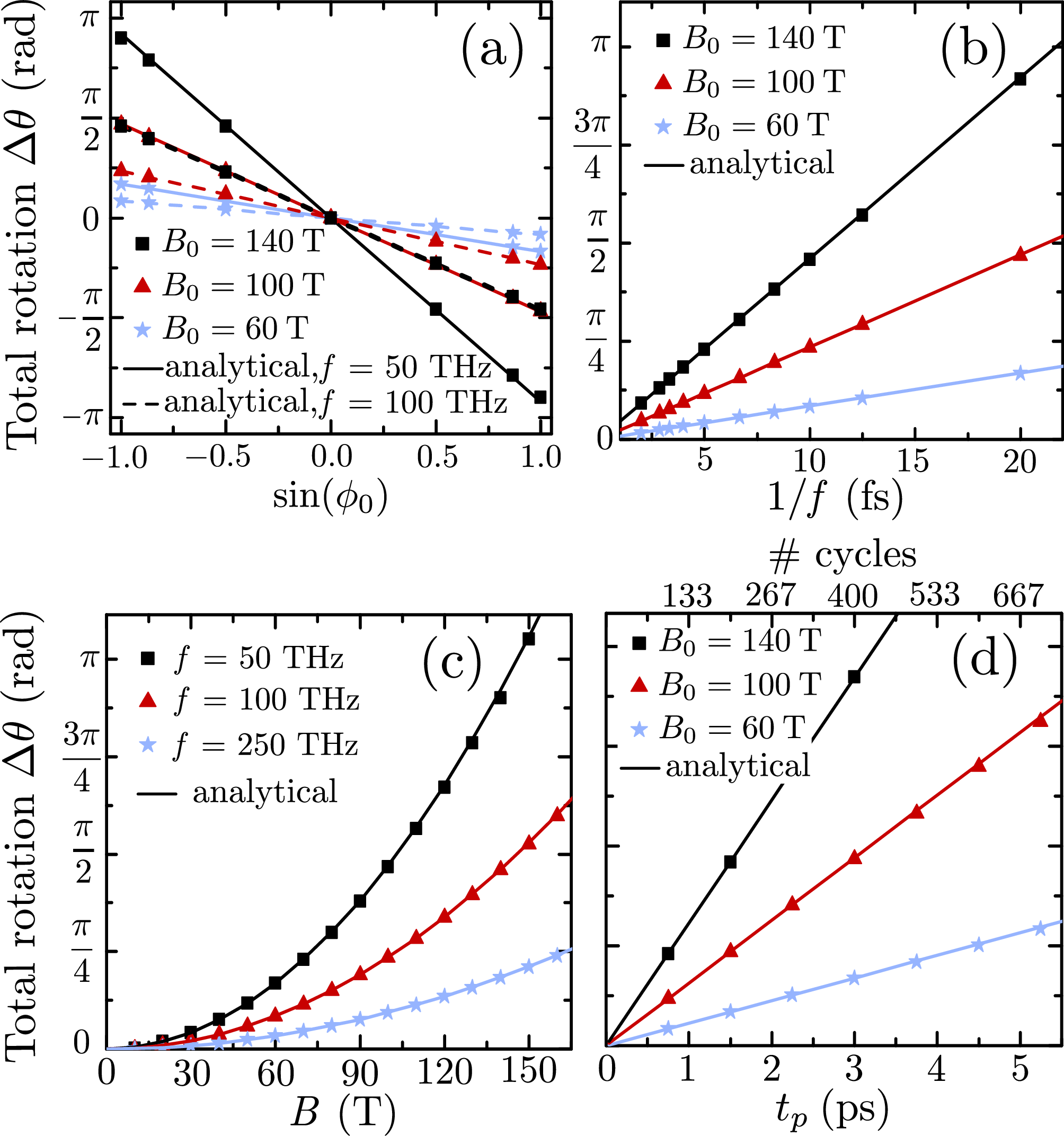}
\caption{%\label{fig:2}
\textbf{Analysis of the nonlinear effect dependencies} Total magnetization rotation as a function of \textbf{(A)} the polarization state of the B-field (characterized by $\phi_0$, and using $\theta_0=\pi/4$), and \textbf{(B)} the inverse of the frequency of a circularly polarized B-field. In both, \textbf{(A)} and \textbf{(B)} three different B-field amplitudes (60 T blue, 100 T red and 140 T black) oscillating at $f=50\;\mathrm{THz}$ are represented. \textbf{(C)} Total magnetization rotation as a function of the circularly polarized B-field amplitude, with three different central frequencies ($f=50\;\mathrm{THz}$ blue, $f=100\;\mathrm{THz}$ red and $f=250\;\mathrm{THz}$ black). In \textbf{(A)}, \textbf{(B)} and \textbf{(C)} the B-field pulse duration is $t_p=3\;\mathrm{ps}$. \textbf{(D)} Total magnetization rotation as a function of the circularly polarized B-field pulse duration, $t_p$, with three different B-field amplitudes ($60\;\mathrm{T}$ blue, $100\;\mathrm{T}$ red and $140\;\mathrm{T}$ black) and a central frequency of $f=50\;\mathrm{THz}$. In all panels symbols indicate results from micromagnetic simulations while lines correspond to Eq. (\ref{eq:maxrot}).}
\end{center}
\end{figure}

One of the most appealing opportunities of this non-linear effect is the possibility to achieve non-thermal ultrafast all-optical switching driven solely by an external circularly polarized B-field. Based on the dependencies presented in Fig. 3, we show in Fig. 4 two different micromagnetic simulation results in which switching is achieved through the use of a RCP B-field pulse. The B-field envelopes of each case are represented in dashed-red lines, whereas the magnetization components $m_x$, $m_y$, $m_z$ are represented in blue, yellow and black, respectively. The first case makes use of a short, 1 ps, 60 THz, 275 T B-field pulse, whereas the second case uses a 10 ps B-field pulse of 60 T and 30 THz. In both cases the $m_z$ component reverses its direction along the course of the pulse, showing that complete switching at the fs or ps timescale can be achieved, depending on the strength, pulse duration and frequency of the B-field. 

\begin{figure}
\begin{center}
\includegraphics[width=\columnwidth]{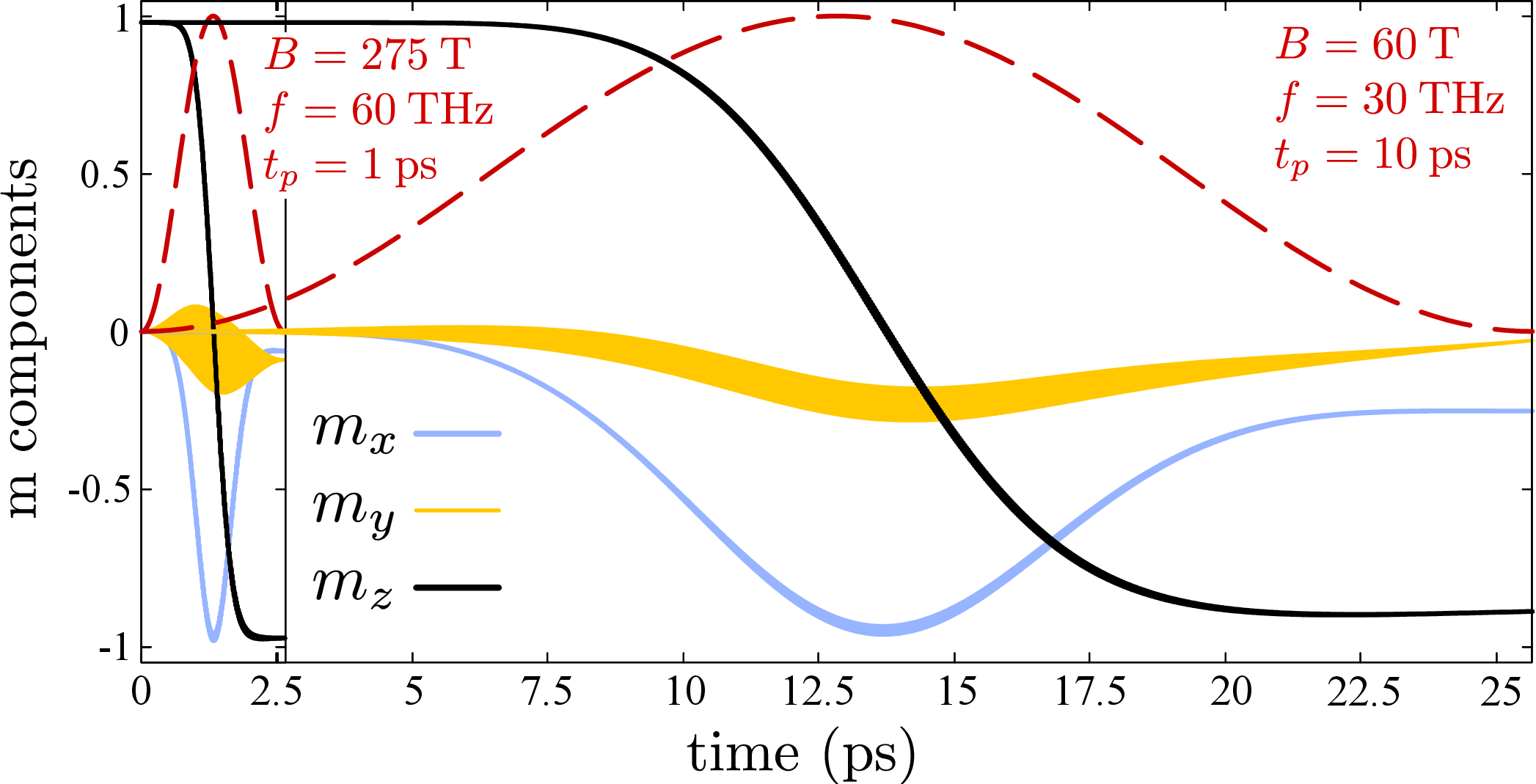}
\caption{%\label{fig:3}
\textbf{Micromagnetic simulation results of the temporal evolution of the magnetization components ($m_x$ blue, $m_y$ yellow, $m_z$ black) of CoFeB excited by a RCP B-field.} The normalized B-field envelope is shown in dashed-red. While a B-field of $B_0=60\;\mathrm{T}$, $f=30$ THz, and $t_p=10$ ps shows switching at the ps timescale, a B-field of $B_0=275\;\mathrm{T}$, $f=60$ THz, and $t_p=1$ ps achieves it at the femtosecond timescale.}
\end{center}
\end{figure}

\section{Discussion}\label{sec:5}
Our results unveil a non-linear chiral magnetic effect driven by ultrafast circularly (or elliptically) polarized B-field pulses, lying in the plane containing the initial magnetization. This purely precessional effect is quadratic in the external B-field, and proportional to the inverse of the frequency, being equivalent to a drift field that depends linearly on the gyromagnetic ratio. This non-linearity is proved to be essential at this time scale, since a linear response would follow adiabatically the magnetic field and, consequently, would restore the magnetization to its initial state after pulse is gone. Conversely, the reported drift field plays a significant role in the magnetization dynamics driven by moderately intense circularly-polarized B-fields ---tens of Tesla at the ps timescale, while hundreds of Tesla at the fs timescale. Although we have studied the magnetization dynamics in CoFeB, this effect is a general feature of the LLG equation, thus being present in all ferromagnets, but also in ferrimagnets and antiferromagnets. Besides, this rectification effect may be exploited to generate THz electric currents via the inverse spin Hall effect, that would emit electromagnetic THz radiation \cite{Seifert:2022} when illuminated with infrared light.

In addition, it should be stressed that, even when the E-field is non-negligible, the reported non-linear mechanism on the B-field may play a role, so a complete study of the ultrafast magnetization dynamics would require taking into account this effect. We note that recent works pointed out the need of including nutation in the dynamical equation of the magnetization \cite{Ciornei:11,Wegrowe:2012,Neeraj:2021}. This term could also lead to second-order effects. Thus, our work serves as a first step towards the investigation of higher-order phenomena induced by magnetic inertia, potentially leading to even shorter time-scale magnetization switching.

Finally, our work demonstrates that the recently developed scenario of spatially isolated fs B-fields \cite{Blanco:2019, Sederberg:2020,Sederberg:2020b,Jana:2021} opens the path to the ultrafast manipulation of magnetization dynamics by purely precessional effects, avoiding thermal effects due to the E-field or magnetization damping. Although the spatial decoupling of the intense B-field from the E-field using fs structured pulses is technologically challenging, it is granted by the rapid development of intense ultrafast laser sources, from the infrared (800 nm, 375 THz), to the mid-infrared (4$\mu$m - 40 $\mu$m, 75 - 7 THz) \cite{Shumakova:2016vj,Gollner2021, Elu:2021wra}. Thinking forward, we believe that our work paves the way towards induced all-optical magnetization dynamics at even shorter timescales, towards the sub-femtosecond regime. Recent works in the generation of ultrafast structured pulses in high-order harmonic generation \cite{HernandezGarcia:2017, HernandezGarcia:2017b, delasHeras:2022} may open the route towards such ultrafast control.\\

\section*{Acknowledgments}
This work was supported by the European Research Council (ERC) under the European Union’s Horizon 2020 research and innovation program (Grant Agreement No. 851201, ATTOSTRUCTURA); Ministerio de Ciencia de Innovación y Universidades (PID2020-117024GB-C41, PID2019-106910GB-I00, RYC-2017-22745); Junta de Castilla y León FEDER (SA287P18).

\bibliographystyle{unsrt}
\bibliography{References}

\end{document}